\documentclass[11pt,twoside,a4paper]{article}

\pdfoutput=1

\usepackage{hyperref}
\usepackage{amsmath}
\usepackage{amssymb}
\usepackage[pdftex]{color,graphicx}
\usepackage[english]{babel}
\usepackage[latin9]{inputenc}
\usepackage[left=3cm,right=3cm,top=3cm,bottom=4cm]{geometry}
\usepackage{color}
\usepackage{multirow}
\usepackage{enumerate}
\usepackage{mdwlist}
\usepackage{verbatim}
\usepackage{color}

\newcommand{\xunit}[1]{\mathrm{#1}}
\newcommand{\unit}[1]{\; \mathrm{#1}}

\newcommand{\MeV}{\unit{MeV}}
\newcommand{\GeV}{\unit{GeV}}

\newcommand{\xMeV}{\xunit{MeV}}
\newcommand{\xGeV}{\xunit{GeV}}

\date{\today}

\begin{document}

\title{
\begin{flushright}
\normalsize{DESY 10-069}
\end{flushright}
\vskip 2cm
{\huge\bf Constraints on a very light CP--odd Higgs of the NMSSM
and other axion--like particles}\\[1cm]}

\author{{\bf\normalsize
Sarah~Andreas, Oleg~Lebedev,
Sa\'ul~Ramos-S\'anchez and Andreas~Ringwald}\\[1cm]
{\it\normalsize
Deutsches Elektronen-Synchrotron DESY}\\
{\it\normalsize Notkestrasse 85, D-22603 Hamburg, Germany
}%\\[20mm]
}

\maketitle

\thispagestyle{empty}

\begin{abstract}
In the NMSSM, a light CP--odd Higgs arises due to spontaneous
breaking of approximate symmetries such as Peccei--Quinn or
R--symmetry and is motivated by string theory. The case when it
is heavier than two muons is well studied and constrained. We
analyze various meson decay, $g-2$, beam dump and reactor
bounds on the CP--odd Higgs with mass below the muon threshold,
in particular, addressing the question how light a CP--odd
Higgs can be. We find that it has to be heavier than 210~MeV or
have couplings to fermions 4 orders of magnitude below those of
the Standard Model Higgs. Our analysis applies more generally
to couplings of a light pseudoscalar to matter.
\end{abstract}

\newpage

\tableofcontents

%%%%
\section{Introduction}
%%%%

An attractive extension of the minimal supersymmetric Standard
Model (MSSM) is the next--to--minimal supersymmetric Standard
Model (NMSSM). Historically, it has been motivated by the
$\mu$--problem of the MSSM which can be solved by replacing the
$\mu$--parameter with an SM singlet $S$~\cite{Ellis:1988er}.
More recently, it has been suggested that the fine--tuning
problem of the MSSM can be alleviated or removed in the NMSSM
when a light CP--odd Higgs is present in the
spectrum~\cite{Dermisek:2005ar,Dermisek:2006wr}. Although this
scenario is now tightly constrained by the new ALEPH
analysis~\cite{Schael:2010aw} as well as \textsc{BaBar} data on
$\Upsilon (3S)$ decays~\cite{Aubert:2009cka,Aubert:2009cp},
some parameter space remains available~\cite{Dermisek:2010mg}
and can further be probed by
$\eta_b$--decays~\cite{Rashed:2010jp}.

The NMSSM--like models can also be obtained from the heterotic
string~\cite{Lebedev:2009ag}, based on the MSSM constructions
of
Refs.~\cite{Buchmuller:2005jr,Buchmuller:2006ik,Lebedev:2006kn,Lebedev:2008un}.
An interesting feature of these models is that the SM singlet
$S$ is not a singlet under the full gauge symmetry of the
heterotic string ($E_8 \times E_8$). Therefore all
self--interactions
\begin{equation}
 S ~,~ S^2 ~,~ S^3 ~,~ ...
\end{equation}
are forbidden by gauge invariance. They are only allowed after
spontaneous breaking of gauge symmetry and the corresponding
couplings are suppressed by VEVs of the symmetry--breaking
fields. On the other hand, interactions like $S H_1 H_2$ can be
allowed without symmetry breaking. Due to this hierarchy in the
couplings, one obtains specific versions of the NMSSM such as
the Peccei--Quinn (PQ) or decoupling
versions~\cite{Lebedev:2009ag}. In the former case, a
pseudo--Goldstone boson appears in the spectrum at the
electroweak scale. Its mass is generated by small PQ violating
effects and can be much below the GeV scale.

Motivated by these considerations and also by a possible
connection of a light pseudoscalar to dark
matter~\cite{Nomura:2008ru,Bai:2009ka,Hooper:2009gm,Gunion:2005rw},
in this work we study constraints on a very light CP--odd Higgs
and address the question how light a CP--odd Higgs can be. We
analyze bounds from various meson decays, muon $g-2$, beam dump
and reactor experiments for the pseudoscalar mass below the
muon threshold. Our results apply beyond the NMSSM to couplings
of a light pseudoscalar to matter.

%%%%
\section{NMSSM and a light CP--odd Higgs}
%%%%

The NMSSM is the MSSM extended by a singlet superfield $S$. In
what follows, we will focus on a particular version of the
NMSSM which has no direct $\mu$--term, the so called
$Z_3$--symmetric NMSSM. The relevant superpotential of the
$Z_3$--symmetric NMSSM is
\begin{equation}
W= \lambda S H_1 H_2 + {1\over 3} \kappa S^3 \;,
\end{equation}
while the soft terms are given by
\begin{equation}
V_{\rm soft}= m_1^2 \vert H_1 \vert^2 + m_2^2 \vert H_2 \vert^2 +
m_S^2 \vert S \vert^2 + \Bigl( \lambda A_\lambda S H_1 H_2 +
{1\over 3} \kappa A_\kappa S^3 + {\rm h.c.}
\Bigr) \;.
\end{equation}
A light pseudoscalar $A^0$ appears naturally in two limiting
cases: when the Higgs potential possesses either approximate
Peccei--Quinn (PQ) or approximate
R--symmetry~\cite{Dobrescu:2000yn,Miller:2003ay,Hall:2004qd,Schuster:2005py,Dermisek:2006wr,Barbieri:2007tu,He:2006fr}.

\subsection{Peccei--Quinn limit}

In the limit $\kappa \rightarrow 0$, the Lagrangian is
invariant under the transformation
\begin{equation}
H_{1,2} \rightarrow e^{i \alpha} H_{1,2} ~~,~~
S \rightarrow e^{-2 i \alpha} S \;.
\end{equation}
At the electroweak scale this symmetry gets broken, resulting
in the ``axion''~\cite{Ellwanger:2009dp}
\begin{eqnarray}
A^0 &=& {1\over N} \Bigl( v~ \sin 2\beta ~ A_{\rm MSSM}^0 - 2s~ S_I \Bigr)
\;, \nonumber\\
N &=& \sqrt{v^2 \sin^2 2\beta + 4 s^2 } \;, \label{axion}
\end{eqnarray}
where $A_{\rm MSSM}^0 = \cos \beta ~H_{1I} + \sin \beta ~
H_{2I}$ is the MSSM pseudoscalar, $s = \langle S \rangle$ and
the subscript $I$ refers to the imaginary part of the Higgs
neutral component. As usual, $\tan\beta = v_1/ v_2$ and $v =
\sqrt{v_1^2 + v_2^2}= 174$ GeV. The mass of the pseudoscalar is
most easily expressed in the large $\tan\beta$
regime~\cite{Miller:2003ay}:
\begin{equation}
m_{A^0}^2 \simeq -3 \kappa A_\kappa s \;. \label{Amass}
\end{equation}
Since renormalization of $\kappa$ is proportional to $\kappa$
itself, this coupling can be very small and $A^0$ very light.
In the string NMSSM example of Ref.~\cite{Lebedev:2009ag},
$\kappa < {\cal O}((\phi/M_{\rm Pl})^5)$ with $\phi$ being an
average VEV of certain SM singlets. For $\phi$ an order of
magnitude below the Planck scale, $\kappa $ can be as small as
$10^{-6}$ leading to a 100 MeV pseudoscalar. In other models it
can be even lighter. Since the PQ symmetry is anomalous as in
the DFSZ construction~\cite{Dine:1981rt}, the lower limit on
$m_{A^0}$ is set by the anomaly contribution of order 100
keV~\cite{Dine:1981rt} (for $s \sim v$).

\subsection{R--symmetry limit}

In the limit $A_\kappa, A_\lambda \rightarrow 0$, the Higgs
sector of the NMSSM is R--invariant. Under R--symmetry the
superfields transform as
\begin{equation}
H_{1,2} \rightarrow e^{i \alpha_R} H_{1,2} ~~,~~
S \rightarrow e^{ i \alpha_R} S \;,
\end{equation}
such that the superpotential transforms with charge 2.
Spontaneous breaking of this symmetry results in an
``R--axion''. Its composition is given
by~\cite{Ellwanger:2009dp}
\begin{eqnarray}
A^0 &=& {1\over N} \Bigl( v~ \sin 2\beta ~ A_{\rm MSSM}^0 + s~ S_I \Bigr)
\;, \nonumber\\
N &=& \sqrt{v^2 \sin^2 2\beta + s^2 } \;, \label{Raxion}
\end{eqnarray}
with $A_{\rm MSSM}^0$ as in Eq.~\ref{axion}.

Unlike the Peccei-Quinn symmetry, R--symmetry is not a
(classical) symmetry of the full Lagrangian. Even if $A_\kappa
, A_\lambda \rightarrow 0$, the gaugino mass terms break it
explicitly. Non--zero A--terms are induced by renormalization,
so their minimal value is a loop factor times the gaugino mass.
The axion mass is again approximated by (\ref{Amass}).

In both PQ-- and R--symmetric cases, the light pseudoscalar is
mostly a singlet in the limit $s \gg v~\sin 2\beta $. Its
couplings to gauge bosons and SM matter are suppressed in this
limit. However, $s$ cannot be too large, otherwise a large
effective $\mu$--term is induced. An exception is the case
$\lambda \ll 1$, which corresponds to the ``decoupling limit'',
i.e. no communication between the singlet and the rest of the
NMSSM.

%%%%
\section{Constraints on a light CP--odd Higgs}\label{sec-lab}
%%%%
Following the notation of Ref.~\cite{Dermisek:2010mg}, the
coupling of the CP-odd Higgs $A^0$ to fermions is given
by
\begin{eqnarray}
\Delta\mathcal{L} &=& - i \frac{ g}{2 m_W} \ C_{Aff} \ \biggl(m_d~ \bar{d}
\gamma_5 d + \frac{1}{\tan^2 \beta} m_u~ \bar{u} \gamma_5 u + m_l~ \bar{l}
\gamma_5 l \biggr) \ A^0 \label{eq-L}.
\end{eqnarray}
In the NMSSM, the coupling $C_{Aff}$ can be expressed in terms
of the singlet--doublet mixing angle $\theta_A$ and
$\tan\beta$: $C_{Aff}= \cos\theta_A
\tan\beta$~\cite{Dermisek:2010mg}, with $\cos\theta_A= v~ \sin
2\beta /N$ and $N$ given in Eqs.~\ref{axion},\ref{Raxion}. In
what follows, we treat it as a free parameter and derive
various particle physics constraints on it. In the NMSSM, very
large ($>10^2$) and very small ($<10^{-2}$) values of $C_{Aff}$
lead to violation of perturbativity and/or finetuning, so it
usually suffices to focus on the moderate $C_{Aff}$ window.
However, our analysis applies more generally to the coupling of
any pseudoscalar to matter as long as the $C_{Aff}$ parameter
is universal for all fermions.

For $m_{A^0} > 2 m_\mu$, the resulting constraints, in
particular from meson decays, are well studied. In the range $2
m_\mu < m_{A^0} < 3 m_\pi$, $A^0$ decays predominantly into 2
muons, which is constrained by $K^+ \rightarrow \pi^+ A^0$ and
$B \rightarrow K~ A^0$. The resulting bound is of order $
C_{Aff} < {\cal O} (10^{-2})$~\cite{Hiller:2004ii}. Above the
3--pion threshold, the branching ratio for the decays into
muons reduces and the bound weakens somewhat. For even larger
$A^0$ mass, the $\Upsilon \rightarrow \gamma A^0$ decay imposes
$C_{Aff} < 0.5$~\cite{Dermisek:2010mg} at $\tan\beta \sim 1$
with the bound getting weaker, ${\cal O}(1)$, close to
$m_\Upsilon$. Above 12 GeV, the DELPHI data on $e^+ e^-
\rightarrow b \bar b A^0 \rightarrow b \bar b b \bar b $ set a
limit $C_{Aff}< {\cal O}(10) $~\cite{Dermisek:2010mg}. Further
constraints, usually relevant at large $\tan\beta$, are
summarized in Ref.~\cite{Ellwanger:2009dp}.

The $m_{A^0} < 2 m_\mu$ territory is less well explored. Some
constraints have been studied in
Refs.~\cite{Hiller:2004ii,Hall:2004qd} in the framework of the
NMSSM, and in Refs.~\cite{Anderson:2003bj,Larios:2001ma} for 2
Higgs Doublet Models (see also~\cite{Dobrescu:1999gv}). In what
follows, we delineate the \{$m_{A^0}, C_{Aff}$\} parameter
space taking into account meson decay, muon $g-2$, nuclear
reactor and beam dump constraints. In particular, we address
the question how light a CP--odd Higgs boson can be. Since we
work in terms of the coupling $C_{Aff}$, most of our results
are largely independent of $\tan\beta$.

For $m_{A^0}$ below the muon threshold, $A^0$ can only decay
into electron-positron pairs and photons. Its total decay width
is
\begin{equation}
\Gamma_{\mathrm{tot}} =
\Gamma (A^0\rightarrow e^+ e^-) + \Gamma (A^0 \rightarrow \gamma \gamma) \label{gamma-total}
\end{equation}
with
\begin{eqnarray}
\Gamma (A^0 \rightarrow f \bar{f}) &=& \frac{\sqrt{2} G_F}{8 \pi}
\ m_f^2 \ m_{A^0} \ C_{Aff}^2 \ \sqrt{1-4 \frac{m_f^2}{m_{A^0}^2}} \label{eq-GammaA0toff} \;, \\
\Gamma (A^0 \rightarrow \gamma \gamma) &=& \frac{\sqrt{2} G_F \alpha^2}{16 \pi^3}
\ m_{A^0}^3 \ \Big|\sum_i r C_{Aii} Q_i^2 k_i F(k_i)\Big|^2 \label{eq-GammaA0togg} \;,
\end{eqnarray}
where $r=1 (N_c)$ for leptons (quarks), $k_i =
m_i^2/m_{A^0}^2$, $Q_i$ is the charge of the fermion in the
loop; $C_{Aii}= C_{Aff}$ for the down--type fermions and
$C_{Aii}= C_{Aff}/\tan^2\beta$ for the up--type fermions. (Here
we neglect the chargino contribution as the coupling is
dominated by the SM fermions.) The loop function $F(k_i)$ is
given by~\cite{Gunion:1988mf}
\begin{equation}
F(k_i) = \left\{
\begin{array}{l l}
-2 (\arcsin \frac{1}{2 \sqrt{k_i}} )^2 & \mathrm{for} \ k_i \geq \frac{1}{4},\\
\frac{1}{2} [ \ln(\frac{1+\sqrt{1-4 k_i}}{1-\sqrt{1-4 k_i}}) + i \pi ]^2 & \mathrm{for} \ k_i < \frac{1}{4},
\end{array} \right.
\end{equation}
and has the limits
\begin{equation}
k_i F(k_i) = \left\{
\begin{array}{r l}
0 & \mathrm{for} \ k_i \ll 1,\\
- \frac{\pi^2}{8} & \mathrm{for} \ k_i = \frac{1}{4}, \\
- \frac{1}{2} & \mathrm{for} \ k_i \gg 1.
\end{array} \right.
\end{equation}
For example, with $m_{A^0}$ = 0.5 MeV and $C_{Aff}$ = 1 the
total width is about $4\times 10^{-12}$ eV. This corresponds to
the decay length (for a boost factor $\gamma\sim 1$) $\tau c
\sim 60$ km. Above the electron threshold, taking $m_{A^0}$ =
50 MeV and $C_{Aff}$ = 1, the total width is $10^{-5}$ eV and
the corresponding decay length is $\tau c \sim 2$ cm.

\begin{figure}[htb!]
\begin{center}
\includegraphics[width=9cm]{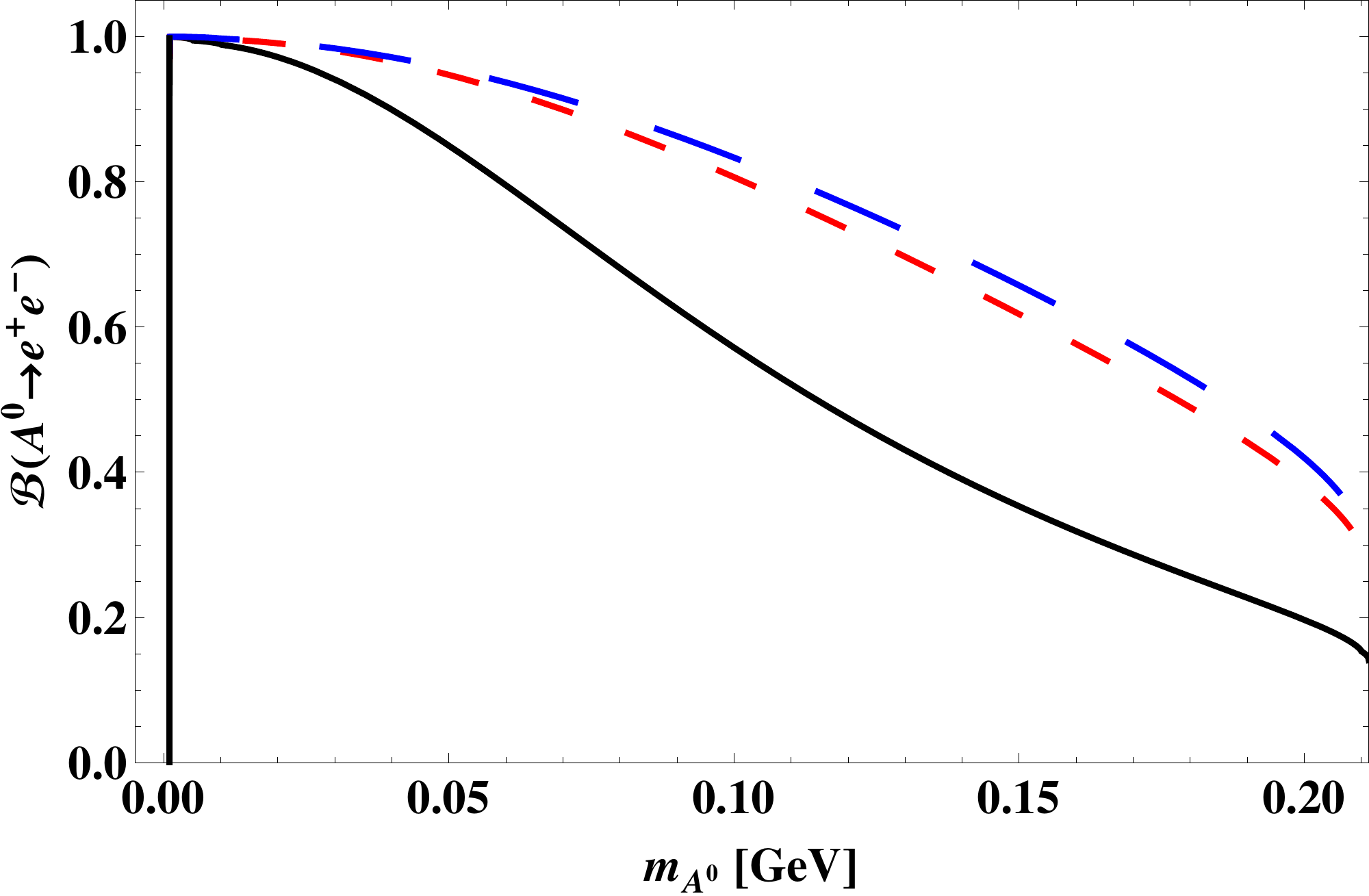}
\caption{$\tan\beta $ dependence of BR$(A^0 \rightarrow e^+ e^-)$. The lowest curve
corresponds to $\tan\beta=1$, the higher one to $\tan\beta=3$, and
the highest curve to $\tan\beta=10$. \label{BR}}
\end{center}
\end{figure}

The $\tan\beta$ dependence of BR$(A^0 \rightarrow e^+ e^-)$ is
shown in Fig.~\ref{BR}. By increasing $\tan\beta$ one reduces
the up--type quark contributions to $\Gamma(A^0 \rightarrow
\gamma\gamma)$, thereby increasing BR$(A^0 \rightarrow e^+
e^-)$. This dependence saturates for $\tan\beta \geq 3$. The
decay mode $A^0 \rightarrow e^+ e^-$ dominates for $m_{A^0}$
below 80 MeV and switches off abruptly just above the electron
threshold.

Below we study various constraints on \{$m_{A^0}, C_{Aff}$\}.
An important class of constraints is due to the decays $$X
\rightarrow Y + {\rm invisible}$$ with $X,Y$ being some mesons.
Experimental limits on their branching ratios exclude parts of
parameter space where $A^0$ is sufficiently long-lived to
escape the detector, that is $\tau \gamma c > d$ with $\tau$
being the lifetime of $A^0$, $\gamma$ a boost factor and $d$
the size of the detector ($\sim 10$ m). In other words,
\begin{eqnarray}
\Gamma_{\mathrm{tot}} & < & \frac{E_{A^0}}{m_{A^0} \ d}. \label{invisible-bound}
\end{eqnarray}
For a two-body decay $X \rightarrow Y+ A^0$,
\begin{equation}
E_{A^0} = \frac{m_X^2 - m_Y^2 + m_{A^0}^2}{2 m_X} \label{eq-EA}
\end{equation}
and in relevant cases $m_X^2 \gg m_Y^2, m_{A^0}^2 $. Then for
$m_{A^0}< 2 m_e$, one has
\begin{eqnarray}
m_{A^0} \sqrt{C_{Aff}}
& \lesssim & 25 \MeV \times \sqrt[4]{\frac{m_X}{\xGeV}} \label{eq-InvAlessme}
\end{eqnarray}
for $d\sim 10$ m and $\tan\beta \sim 1$. For $m_{A^0}> 2 m_e$,
one can estimate the resulting bound by taking
$\Gamma_{\mathrm{tot}} \sim \Gamma_{e^+ e^-}$, in which case
\begin{eqnarray}
m_{A^0} C_{Aff} & \lesssim & 8 \MeV \times \sqrt{\frac{m_X}{\xGeV}} \ .
\label{eq-InvAgtrme}
\end{eqnarray}

%%%%%%%%%%%%%%%%%%%%%%%%%%%%%%%%%%%%%%%%%%%%%%%%%%%%%%%%%%%%%%%%%%%%%%%%%%%%%%%      invisible B-decay      %%%%%%%%%%%%%%%

%%%%
\subsection{Rare B--decays $B \rightarrow K + {\rm invisible}$} \label{sec-B}
%%%%

Limits on production of $A^0$ in rare $B$-meson
decays~\cite{Hall:1981bc,Frere:1981cc,Freytsis:2009ct,Batell:2009jf}
result from the following bounds on the branching ratios
measured by \textsc{CLEO}~\cite{Ammar:2001gi} and
\textsc{BaBar}~\cite{Aubert:2003yh}:
\begin{eqnarray}
& \mathcal{B}^{\textsc{CLEO}}(B^0 \rightarrow K_S^0 + {\rm invisible}) & < ~ 5.3 \times 10^{-5} \;,\\
& \mathcal{B}^{\textsc{BaBar}}(B^- \rightarrow K^- \nu \bar{\nu}) & < ~ 7.0 \times 10^{-5} \;,\nonumber
\end{eqnarray}
where in what follows we use the more constraining
\textsc{CLEO} result. Since in both experiments $A^0$ appears
as missing energy, the bounds apply only if $A^0$ decays
outside the detector. For $m_X = m_{B^0} = 5.28 \GeV$, this
implies
\begin{eqnarray}
m_{A^0} \sqrt{C_{Aff}} ~~ \lesssim & 37 & \xMeV ~~~ \mathrm{for} \ m_{A^0} < 2m_e \;,\nonumber \\
m_{A^0} \ C_{Aff} ~~ \lesssim & 18 & \xMeV ~~~ \mathrm{for} \ m_{A^0} > 2m_e \; . \label{eq-InvAfromB}
\end{eqnarray}

In the NMSSM, the decay rate for $B^0 \rightarrow K^0 A^0$
is given by~\cite{Hiller:2004ii}
\begin{equation}
\Gamma (B^0 \rightarrow K^0 A^0) = \frac{G_F^2 \ |V_{tb} V_{ts}^\ast|^2}{2^{10} \pi^5} \
|C_A|^2 \ \frac{|\vec{p}_K|}{m_{B^0}^2} \ \Bigl\vert f_0^{B^0} (m_{A^0}^2)\Bigr\vert^2 \
\Big( \frac{m_{B^0}^2 - m_{K^0}^2}{m_b} \Big)^2, \label{eq-GammaBtoK}
\end{equation}
where the form factor $f_0^{B^0}(0) \sim
0.3-0.4$~\cite{Ali:1999mm} and $\vert \vec{p}_K \vert \simeq
m_{B^0}/2$ is the three momentum of the kaon. The quantity
$C_A$ has been calculated in Ref.~\cite{Hiller:2004ii} in the
large $\tan\beta$ regime, $C_A \sim C_{Aff} \tan \beta ~m_b
m_t$ for order one stop mixing and EW scale sparticles. Since
the full NMSSM calculation at low $\tan\beta$ is not available,
we estimate the order of magnitude of the resulting bound by a
rescaling of this result. Using the total $B^0$ width
$\Gamma_{B^0} = 4.3 \times 10^{-13} \GeV$, the CLEO bound
implies $C_{Aff}< 0.02/\tan\beta$. Taking conservatively
$\tan\beta \sim {\cal O} (1)$, we get{\footnote{Essentially,
this corresponds to the SM contribution with an additional
coupling (\ref{eq-L}). The $b_R-s_L$ transition is mediated by
the $W-t$ loop with $A^0$ coupled to the top quark. }}
\begin{equation}
C_{Aff} < 10^{-2} \;.
\end{equation}
This constraint is already strong at small $\tan\beta$ and gets
even stronger at large $\tan\beta$.

The resulting exclusion region is shown in
Fig.~\ref{fig-ConstraintsDecays} (marked $``B^0 \rightarrow K^0
+ {\rm inv.}"$). We note that, in contrast to the lower boundary, the right
boundary of this region is calculated quite reliably from
Eq.~\ref{invisible-bound} and is essentially independent of
$\tan\beta$. In the plot, we use the full $A^0$-width
$\Gamma_{\mathrm{tot}}$ without resorting to the approximation
$ \Gamma_{\mathrm{tot}} \sim \Gamma_{e^+ e^-}$. The kink at
$m_{A^0}\approx 2 m_e$ is due to the rapid fall of $\Gamma(A^0
\rightarrow e^+ e^-)$ as $m_{A^0}$ approaches the threshold
from above. Finally, the dependence on the detector size is
only square--root.

%%%%%%%%%%%%%%%%%%%%%%%%%%%%%%%%%%%%%%%%%%%%%%%%%%%%%%%%%%%%%%%%%%%%%%%%%%%%%%%      invisible K-decay      %%%%%%%%%%%%%%%

%%%%
\subsection{Rare K--decays $K \rightarrow \pi + {\rm invisible} $ } \label{sec-K}
%%%%

\begin{figure}[htb!]
\begin{center}
\includegraphics[width=13cm]{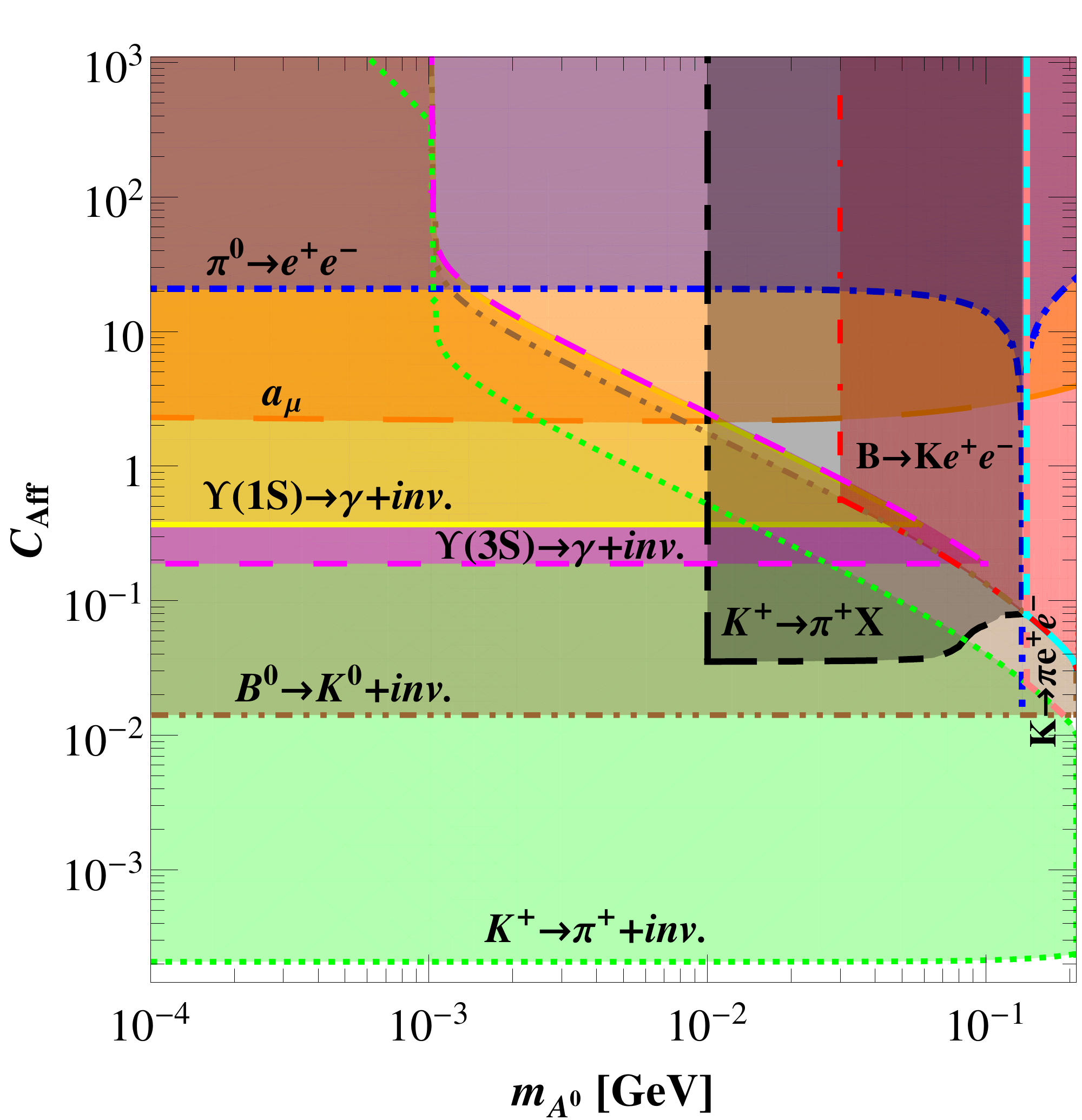}
\caption{Constraints from meson decays and muon $g-2$. The colored
regions are excluded.
These bounds include the effect of varying BR$(A^0 \rightarrow e^+ e^-)$
with $m_{A^0}$.
\label{fig-ConstraintsDecays}}
\end{center}
\end{figure}

A light (invisible) $A^0$ can also be produced in $K$--decays.
The relevant branching ratio has been measured by
\textsc{E787}~\cite{Adler:2004hp,Adler:2002hy} and
\textsc{E949}~\cite{Artamonov:2009sz}:\footnote{These bounds
become significantly weaker at the pion pole, $m_{A^0}=m_\pi$.}
\begin{eqnarray}
\mathcal{B}^{\textsc{E787}}(K^+ \rightarrow \pi^+ + \mathrm{invisible}) & < & 4.5 \times 10^{-11}, \\
\mathcal{B}^{\textsc{E949}}(K^+ \rightarrow \pi^+ + \mathrm{invisible}) & < & 10^{-10}. \nonumber
\end{eqnarray}
We will use the tighter E787 bound. Eqs.~\ref{eq-InvAlessme}
and~\ref{eq-InvAgtrme} for $m_X = m_{K^+} = 494 \MeV$ yield the
``invisibility'' conditions
\begin{eqnarray}
m_{A^0} \sqrt{C_{Aff}} ~~ \lesssim & 21 & \xMeV ~~~ \mathrm{for} \ m_{A^0} < 2 m_e \;, \nonumber \\
m_{A^0} \ C_{Aff} ~~ \lesssim & 5 & \xMeV ~~~ \mathrm{for} \ m_{A^0} > 2 m_e \;. \label{eq-InvAfromK}
\end{eqnarray}
The decay rate is given by
\begin{equation}
\Gamma (K^+ \rightarrow \pi^+ A^0) = \frac{G_F^2 \ |V_{ts} V_{td}^\ast|^2}{2^{10} \pi^5} \
|C_A'|^2 \ \frac{|\vec{p}_\pi|}{m_{K^+}^2} \ \Bigl\vert f_0^{K^+} (m_{A^0}^2)\Bigr\vert^2 \
\Big( \frac{m_{K^+}^2 - m_{\pi^+}^2}{m_s} \Big)^2 \;, \label{eq-GammaKtoPi}
\end{equation}
where the form factor $f_0^{K^+}(0) \sim
1$~\cite{Marciano:1996wy}, $\vert \vec{p}_\pi \vert \simeq
m_{K^+}/2$ and $C_A' \sim C_{Aff} \tan \beta ~m_s m_t$. With
$\Gamma_{K^+}= 5.32 \times 10^{-17} \GeV $, the experimental
bound requires $C_{Aff}< 2 \times 10^{-4}/\tan\beta $, which
for $\tan\beta \sim {\cal O}(1)$ gives
\begin{equation}
C_{Aff} < 10^{-4} \;.
\end{equation}
The corresponding excluded region is marked ``$K^+ \rightarrow
\pi^+ + {\rm inv.}$'' in Fig.~\ref{fig-ConstraintsDecays}. As
in the case of $B$--decays, this bound only gets stronger with
increasing $\tan\beta$ and its precise value is not important
for us.

%%%%%%%%%%%%%%%%%%%%%%%%%%%%%%%%%%%%%%%%%%%%%%%%%%%%%%%%%%%%%%%%%%%%%%%%%%%%%%%      visible B-/K-decays      %%%%%%%%%%%%%%%

\subsection{Rare decays $B \rightarrow K ~e^+e^-$, $K \rightarrow \pi ~e^+e^-$}\label{sec-e+e-}

If $A^0$ decays inside the detector, it contributes to the
processes $B \rightarrow K ~e^+e^-$, $K \rightarrow \pi
~e^+e^-$. In experimental measurements, one sets a cut on the
invariant mass $m_{e^+ e^-} > 140$ MeV in order to suppress
backgrounds including conversion photons and $\pi^0 \rightarrow
e^+ e^- \gamma$. Therefore, the resulting bounds apply for
$m_{A^0} > 140$ MeV.

\textsc{BELLE} has reported~\cite{Ishikawa:2003cp}
\begin{equation}
\mathcal{B}^\textsc{BELLE} (B \rightarrow K ~\ell^+ \ell^- ) =
(4.8^{+1.0}_{-0.9} \pm 0.3 \pm 0.1) \times 10^{-7} \;,
\end{equation}
where $\ell$ includes muons and electrons with $m_{e^+ e^-} >
140$ MeV. Assuming lepton universality, we will use a
conservative bound BR$(B \xrightarrow{A^0} K ~e^+ e^- ) < 2.4
\times 10^{-7}$. The NMSSM result can be read off from
Eq.~\ref{eq-GammaBtoK} using
\begin{equation}
{\rm BR}(B \xrightarrow{A^0} K ~e^+ e^- ) \simeq {\rm BR}(B \rightarrow K ~A^0 ) \times
{\rm BR } (A^0 \rightarrow e^+ e^-) \;. \label{Kee}
\end{equation}
In the relevant mass range, ${\rm BR } (A^0 \rightarrow e^+ e^-)$ is 20 - 40 \% for
$\tan\beta \sim 1$.
After imposing the condition that $A^0$ decay inside the detector,
we get
\begin{equation}
C_{Aff} < 8 \times 10^{-2} \;,
\end{equation}
which excludes the strip $140 \MeV < m_{A^0} < 2 m_\mu$ shown
in Fig.~\ref{fig-ConstraintsDecays} (marked $``B \rightarrow K
e^+ e^- "$).

A similar result is obtained from
$K$--decays~\cite{Goudzovski:2009mg}:
\begin{equation}
\mathcal{B}^\textsc{NA48/2} (K^\pm \rightarrow \pi^\pm ~e^+ e^- ) =
(3.11\pm0.04\pm0.05\pm0.08\pm0.07) \times 10^{-7} \;,
\end{equation}
which also employs the same kinematic cut $m_{e^+ e^-} > 140$ MeV.
Using Eq.~\ref{eq-GammaKtoPi} and an analog of (\ref{Kee}), we get
\begin{equation}
C_{Aff} < 2 \times 10^{-2} \;,
\end{equation}
where the ``visibility'' condition for $A^0$ has been imposed.
This bound is shown in Fig.~\ref{fig-ConstraintsDecays} marked
$``K \rightarrow \pi e^+ e^- "$.

It is noteworthy that the window $140 \MeV < m_{A^0} < 2 m_\mu$
is eliminated by 2 different processes. There are additional
NMSSM contributions to $B \rightarrow K ~e^+e^-$, $K
\rightarrow \pi ~e^+e^-$ apart from that of $A^0$, so in
principle there could be cancellations. Considering two
independent processes makes this possibility less likely.

The \textsc{BaBar} measurement of BR$ (B \rightarrow K ~\ell^+
\ell^- )$~\cite{Aubert:2005cf} imposes a lower kinematic cut
$m_{e^+ e^-} > 30$ MeV thereby losing somewhat in efficiency
because of the low energy backgrounds~\cite{Hiller:2004ii}.
Their result
\begin{equation}
\mathcal{B}^{\textsc{BaBar}} (B \rightarrow K ~\ell^+ \ell^- )=
(0.34 \pm 0.07 \pm 0.03) \times 10^{-6}
\end{equation}
excludes the region $m_{A^0}> 30$ MeV with $C_{Aff}> 10^{-1}-1$
(where $A^0$ decays inside the detector) depending on the
pseudoscalar mass, Fig.~\ref{fig-ConstraintsDecays} (marked
$``B \rightarrow K e^+ e^-"$). Although one may question the
reliability of this result at low $e^+e^-$ invariant masses,
another experiment, to be discussed in the next subsection,
excludes a similar region of parameter space.

%%%%%%%%%%%%%%%%%%%%%%%%%%%%%%%%%%%%%%%%%%%%%%%%%%%%%%%%%%%%%%%%%%%%%%%%%%%%%%%      rare K-decays      %%%%%%%%%%%%%%%

\subsection{Rare K--decays $K^+ \rightarrow \pi^+ + X $ } \label{sec-K-anything}

A byproduct of the $K_{\mu 2}$ experiment in Japan was a
measurement of a 2-body decay $K^+ \rightarrow \pi^+ + X $,
where $X$ is any particle~\cite{Yamazaki:1984vg}. One searched
for a peak in the $\pi^+$ momentum for 10 MeV $< m_{X}<$ 300
MeV. The resulting bound is
\begin{equation}
\mathcal{B}(K^+ \rightarrow \pi^+ + X) < 10^{-6}
\end{equation}
at 90\% CL for $m_X <$ 60 MeV. For larger $m_X$ up to 120 MeV
this bound relaxes to $10^{-5}$.

The excluded parameter space is shown in
Fig.~\ref{fig-ConstraintsDecays} (marked $``K^+ \rightarrow \pi^+
X" $). The constraint amounts approximately to
\begin{equation}
C_{Aff} < 4 \times 10^{-2}
\end{equation}
for $m_{A^0} >$ 10 MeV.

We note that at $m_{A^0}\simeq m_{\pi^0}$ the constraint is
weaker. However, this region is disfavored by $\pi^0
\rightarrow e^+ e^-$ (see below) and the $\pi^+ - \pi^0$ mass
difference when $\pi^0 - A^0$ mixing is taken into account.

A similar region of parameter space (up to $m_{A^0} = 100$ MeV)
is excluded by the process $\pi^+ \rightarrow e^+ \nu ~A^0$
with subsequent decay $A^0 \rightarrow e^+
e^-$~\cite{Eichler:1986nj} (for applications to axion models,
see~\cite{Bardeen:1986yb}).

%%%%%%%%%%%%%%%%%%%%%%%%%%%%%%%%%%%%%%%%%%%%%%%%%%%%%%%%%%%%%%%%%%%%%%%%%%%%%%%      Upsilon decay      %%%%%%%%%%%%%%%

%%%%
\subsection{Radiative Upsilon--decays} \label{sec-Ups}
%%%%

The bounds on $A^0$ production in radiative $\Upsilon$ decays
come from \textsc{CLEO}~\cite{Balest:1994ch} and
\textsc{BaBar}~\cite{Sekula:2008sb,Aubert2008st}:
\begin{eqnarray}
& \mathcal{B}^{\textsc{CLEO}}(\Upsilon(1S) \rightarrow \gamma + \mathrm{invisible} ) & < ~ 1.3 \times 10^{-5}, \\
& \mathcal{B}^{\textsc{BaBar}}(\Upsilon(3S) \rightarrow \gamma + \mathrm{invisible} ) & < ~ 3 \times 10^{-6}.
\end{eqnarray}
They apply only if $A^0$ decays outside the detector, which for
$m_X = m_{\Upsilon(3S)} = 10.4 \GeV$ means
\begin{eqnarray}
m_{A^0} \sqrt{C_{Aff}} ~~ \lesssim & 44 & \xMeV ~~~ \mathrm{for} \ m_{A^0} < 2m_e \;, \nonumber \\
m_{A^0} \ C_{Aff} ~~ \lesssim & 25 & \xMeV ~~~ \mathrm{for} \ m_{A^0} > 2m_e \;. \label{eq-InvAfromU3S}
\end{eqnarray}
The branching ratio for $\Upsilon \rightarrow A^0 \gamma $ is
given by~\cite{Wilczek:1977pj,Haber:1987ua,Hiller:2004ii}
\begin{equation}
\frac{\mathcal{B} (\Upsilon \rightarrow A^0 \gamma)}{\mathcal{B} (\Upsilon \rightarrow \mu^+ \mu^-)} =
\frac{G_F m_b^2}{ \sqrt{2} \pi \alpha} \ C_{Aff}^2 \ \Big(1 - \frac{m_{A^0}^2}{m_\Upsilon^2} \Big) ~F_{\rm QCD} \;,
\end{equation}
where $F_{\rm QCD} \sim 0.5$ is a QCD correction factor,
$\mathcal{B} (\Upsilon(1S) \rightarrow \mu^+ \mu^-) = 0.025$
and $\mathcal{B} (\Upsilon(3S) \rightarrow \mu^+ \mu^-) =
0.022$~\cite{Amsler:2008zzb}. The resulting bounds (marked in
Fig.~\ref{fig-ConstraintsDecays} ``$\Upsilon(1S) \rightarrow
\gamma + {\rm inv.}$'' and ``$\Upsilon(3S) \rightarrow \gamma +
{\rm inv.}$'') are
\begin{eqnarray}
&& C_{Aff} < 0.37 ~~ ({\rm CLEO}) \;, \nonumber\\
&& C_{Aff} < 0.19 ~~ (\textsc{BaBar}) \;,
\end{eqnarray}
independent of $\tan\beta$.

%%%%%%%%%%%%%%%%%%%%%%%%%%%%%%%%%%%%%%%%%%%%%%%%%%%%%%%%%%%%%%%%%%%%%%%%%%%%%%%      Pion decay      %%%%%%%%%%%%%%%

\subsection{Pion decay $\pi^0 \rightarrow e^+ e^-$}\label{sec-pion}

A light $A^0$ provides a pseudoscalar channel for pion
annihilation into $e^+ e^-$ (see, e.g.~\cite{Chang:2008np}).
This chirality--suppressed decay proceeds in the SM through a
loop diagram with a $\pi^0 \gamma \gamma^*$ vertex and has a
very small branching ratio. The recent \textsc{KTeV}
result~\cite{Abouzaid:2006kk}
\begin{equation}
\mathcal{B}^{\textsc{KTeV}}(\pi^0 \rightarrow e^+ e^-) =
(7.48 \pm 0.29 \pm 0.25) \times 10^{-8}
\end{equation}
is somewhat (3$\sigma$) above the SM
prediction~\cite{Dorokhov:2007bd,Dorokhov:2009xs}. To be
conservative, in what follows we will require that the tree
level contribution from $A^0$ not exceed the central
experimental value, $\mathcal{B} (\pi^0 \xrightarrow{A^0} e^+
e^-) < 7.5 \times 10^{-8} $.

We find
\begin{eqnarray}
\Gamma(\pi^0 \xrightarrow{A^0} e^+ e^-) & \simeq & \frac{G_F^2}{4 \pi} \
\frac{m_e^2 m_\pi^5 f_\pi^2}{ \vert m_\pi^2 - m_{A^0}^2 + i \Gamma_{A^0} m_{A^0} \vert ^2} \ C_{Aff}^4 \;,
\end{eqnarray}
where we have neglected the up--quark contribution, $\langle 0
\vert m_d \bar d \gamma^5 d \vert \pi^0 \rangle \simeq -i
m_\pi^2 f_\pi$.{\footnote{Our bound on $C_{Aff}$ is not
sensitive to this approximation as it scales as a square root
of this matrix element.}} $\Gamma_{A^0} $ is given by
Eq.~\ref{gamma-total}, $m_\pi = 135 \MeV$ and $f_\pi = 93
\MeV$. We neglect the $\pi^0 - A^0$ mixing effects which are of
order $\delta m^2/m_{\pi}^2 \sim f_\pi/M_W \sim 10^{-3}$ and
relevant only very close to the pion mass.

The total width of $\pi^0$ is
\begin{equation}
\Gamma(\pi^0 \rightarrow \gamma \gamma) =
\frac{\alpha^2}{64 \pi^3} \ \frac{m_\pi^3}{f_\pi^2}\;,
\end{equation}
then the \textsc{KTeV} result requires
\begin{equation}
C_{Aff} < 20
\end{equation}
away from the resonance region, where the constraint is
stronger. A more precise bound including the
$m_{A^0}$--dependence is shown in
Fig.~\ref{fig-ConstraintsDecays} marked $``\pi^0 \rightarrow
e^+ e^- "$. This constraint is complementary to those of the $X
\rightarrow Y + {\rm invisible}$ decays, in that it excludes
parameter space above $C_{Aff} \simeq 20$ regardless of the
$A^0$ mass. It is also a reliable tree--level constraint
essentially independent of $\tan\beta$.

%%%%%%%%%%%%%%%%%%%%%%%%%%%%%%%%%%%%%%%%%%%%%%%%%%%%%%%%%%%%%%%%%%%%%%%%%      muon anomalous magn. moment      %%%%%%%%%%%%%%%

%%%%
\subsection{Muon anomalous magnetic moment} \label{sec-amu}
%%%%

At loop level, $A^0$ contributes to the muon $g-2$ which is
well measured. Currently, there is a 4$\sigma$ discrepancy
between the SM prediction for the muon $g-2$ and its measured
value at \textsc{BNL E821}~\cite{Teubner:2010ah}:
\begin{equation}
\Delta a_\mu = a_\mu^{\mathrm{Exp}} - a_\mu^{\mathrm{SM}} =
(31.6 \pm 7.9) \times 10^{-10} \;,
\end{equation}
which may be considered a hint for new physics.

In the NMSSM, there are significant one- and two-loop
contributions of the CP-odd Higgs $A^0$ to $a_\mu$. They are
given, for example, in~\cite{Domingo:2008bb}:
\begin{eqnarray}
\delta a_\mu (A^0) & = & \delta a_\mu^{1\mathrm{L}}(A^0) + \delta a_\mu^{2\mathrm{L}}(A^0)\;, \\
\delta a_\mu^{1\mathrm{L}}(A^0) & = & - \frac{\sqrt{2} G_F}{8 \pi^2} \ m_\mu^2 \ |C_{Aff}|^2 \
f_1 \Big( \frac{m_{A^0}^2}{m_\mu^2} \Big)\;, \nonumber \\
\delta a_\mu^{2\mathrm{L}}(A^0) & = & \frac{\sqrt{2} G_F \alpha}{8 \pi^3} \ m_\mu^2 \ |C_{Aff}|^2 \
\Bigg[ \frac{4}{3} \frac{1}{\tan^2\beta} f_{2}
\Big( \frac{m_t^2}{m_{A^0}^2} \Big) + \frac{1}{3} f_{2}
\Big( \frac{m_b^2}{m_{A^0}^2} \Big) + f_{2} \Big( \frac{m_\tau^2}{m_{A^0}^2} \Big) \Bigg]\;, \nonumber
\end{eqnarray}
where
\begin{eqnarray}
f_1 (z) & = & \int_0^1 dx \frac{x^3}{x^2 + z (1-x)}\;, \nonumber \\
f_2 (z) & = & z \int_0^1 dx \frac{1}{x (1-x) - z} \ln \frac{x(1-x)}{z}\;.
\end{eqnarray}
The one loop-contribution is negative which makes the
discrepancy worse. For $m_{A^0}$ above roughly 1 GeV, the
two--loop contribution may be dominant and resolve the
discrepancy. However, this does not occur in the mass range we
consider.

Since there are NMSSM contributions to $g-2$ of both signs, the
contribution from the CP--odd Higgs $A^0$ can be canceled. We
then require that the latter not worsen the discrepancy beyond
$5~\sigma$:
\begin{equation}
\delta a_\mu (A^0) \leq a_\mu^{\mathrm{Exp}} - a_\mu^{\mathrm{SM}} \simeq
40 \times 10^{-10} \ ~~ \ (5\sigma) \;.
\end{equation}
The corresponding bound on $C_{Aff}$ (at $\tan \beta\sim 1$) is
shown in Fig.~\ref{fig-ConstraintsDecays} marked ``$a_\mu$''.
It can be approximated by
\begin{equation}
C_{Aff} < 2 \;
\end{equation}
for $m_{A^0} \lesssim m_{\mu}$. The $\tan \beta$ dependence is
very mild in the region of interest and stems only from the
2--loop contribution, which is subdominant.

Once this bound is imposed, the electron $g-2$ constraint is
satisfied automatically.

%%%%%%%%%%%%%%%%%%%%%%%%%%%%%%%%%%%%%%%%%%%%%%%%%%%%%%%%%%%%%%%%%%%%%%%%%%%%%%%      Other limits      %%%%%%%%%%%%%%%

\subsection{Other constraints}

Further (model--dependent) constraints are summarized in
Refs.~\cite{Ellwanger:2009dp,Hiller:2004ii}. These are weaker
than the bounds we have considered and require assumptions
about the NMSSM spectrum. For instance, there are contributions
from all neutral Higgses to $B_s \rightarrow \mu^+ \mu^-$ and
$B-\bar B $ mixing which allows one to eliminate parts of
parameter space ($C_{Aff}\sim{\cal O}(10)$) depending on their
masses and $\tan\beta$~\cite{Hiller:2004ii}.

Among other possible flavor physics constraints are $J/\Psi$
decays. \textsc{CLEO} has recently reported
$\mathcal{B}^{\textsc{CLEO}}(J/\Psi \rightarrow \gamma +
\mathrm{invisible}) < ~ 4.3 \times
10^{-6}$~\cite{Insler:2010jw}, which is somewhat weaker than
the analogous $\Upsilon(3S)$ bound. The $A^0$ coupling to
up--type quarks falls very quickly with $\tan\beta$, so we do
not use this result in our analysis.

Further, the missing--energy process $B \rightarrow K A^0 A^0$
proceeding through the $h A^0A^0$ coupling~\cite{Bird:2004ts}
sets a mild constraint on the $SH_1 H_2$ coupling in the
superpotential, $\lambda < 0.7$.

A light CP--odd Higgs could potentially be constrained by the
LEP data. $A^0$ couples to the $Z$--boson at tree level through
the $A^0 H_i^0 Z_{\mu}$ vertex~\cite{Ellwanger:2009dp},
therefore the (invisible) $Z$--width does not constrain the
mass of $A^0$. Furthermore, the electroweak oblique corrections
are suppressed by the mass of the heavier pseudoscalar (see,
e.g.~\cite{Bai:2009ka}). For the same reason, the $A^0$
production at LEP through $e^+ e^- \rightarrow h~A^0$ is
suppressed. The associated production with bottom quarks $e^+
e^- \rightarrow b \bar b ~A^0$ is also
insignificant~\cite{Bai:2009ka}. Finally, the constraints from
$Z \rightarrow \gamma ~A^0$ are
weak~\cite{Raychaudhuri:1991rf,Rupak:1995kg}.

Astrophysical bounds have been summarized in
Ref.~\cite{Hall:2004qd}. They are usually relevant for sub--MeV
pseudoscalar masses which, in the range $10^{-4} < C_{Aff} <
10^3$, are already excluded by meson decays with missing
energy. However, the supernova SN1987A sets an additional
constraint for a small coupling: $ C_{Aff}> 10^{-4}$ when
$m_{A^0} < 30$ MeV~\cite{Hall:2004qd}.

To summarize this section, we see that a combination of various
constraints requires the CP--odd Higgs to be heavier than $2
m_\mu$ (unless $ C_{Aff}< 10^{-4}$). To obtain this bound we
did not rely on the specifics of the NMSSM. All we used was the
coupling (\ref{eq-L}) at $\tan\beta \sim {\cal O}(1)$, which is
much more general. This coupling is sufficient to induce the
$b-s$ and $s-d$ transitions (with flavor change due to SM
loops) which were used in the processes like $B \rightarrow
 K ~A^0$ and $K \rightarrow \pi ~A^0$.
Similarly, $\Upsilon$ decays, $\pi^0 \rightarrow e^+ e^-$ and
muon $g-2$ are generated directly by (\ref{eq-L}).

For completeness, in the next section we discuss the reactor
and beam dump results which have been used in the past to
constrain axion models.

%%%%%%%%%%%%%%%%%%%%%%%%%%%%%%%%%%%%%%%%%%%%%%%%%%%%%%%%%%%%%%%%%%%%%%%%%%%%%%%      Reactor limits      %%%%%%%%%%%%%%%

%%%%
\section{Further bounds from reactor and beam dump experiments} \label{sec-Reactor}
%%%%

\subsection{Reactor bounds}

Searches for axion--like particles using nuclear power reactors
set constraints on the parameter space of the CP--odd Higgs.
Here we consider 2 representative experiments which employ
Bugey and Kuo--Sheng nuclear reactors.

\begin{figure}[htb!]
\begin{center}
\includegraphics[width=9cm]{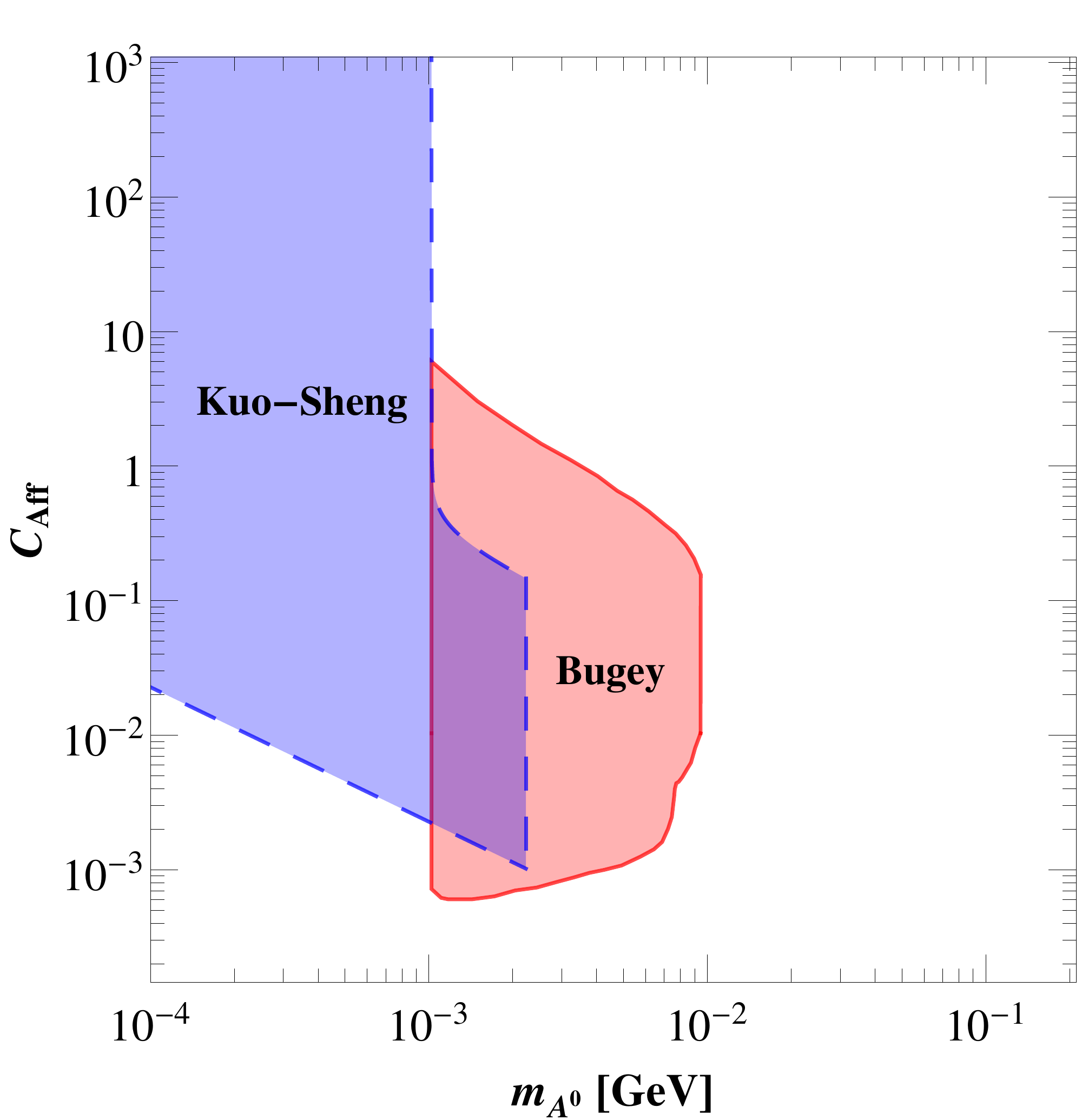}
\caption{Constraints from nuclear power reactors.
\label{fig-ConstraintsReactor}}
\end{center}
\end{figure}

Axion--like particles can be emitted in place of photons from
excited nuclear levels which makes nuclear reactors a source of
pseudoscalars with masses up to 10 MeV.
In~\cite{Altmann:1995bw}, the detector was placed 18.5 m from
the Bugey reactor core and one searched for the decays $A^0
\rightarrow e^+ e^-$. No excess of $e^+ e^-$ events has been
observed which set a constraint on the axion decay constant
$f_\chi$. The corresponding exclusion region can be read off
from Fig.~5 of~\cite{Altmann:1995bw} using the conversion
\begin{equation}
C_{Aff} = \frac{1}{f_\chi} \ \frac{2 m_W}{g} \;. \label{conversion}
\end{equation}
The result is shown in Fig.~\ref{fig-ConstraintsReactor}.

Another experiment at the Kuo--Sheng nuclear reactor searched
for axions via Compton conversion on
electrons~\cite{Chang:2006ug}. A Germanium detector, placed 28
m away from the reactor, measured the ionization energy
resulting from the axion--photon conversion in the detector.
Again, no signal was found. The exclusion region can be read
off from Fig.~7 of~\cite{Chang:2006ug} and their Eq.~31,
$g_{aee}~ g_{a NN}^1 < 1.3 \times 10^{-10}$. Using $g_{aee} =
C_{Aff} g m_e/(2 m_W)$ and $g_{a NN}^1=3 \times 10^{-8}
m_{A^0}/{\rm eV}$, this translates into
\begin{equation}
m_{A^0} \ C_{Aff} < 2 \times 10^{-3}~ {\rm MeV}
\end{equation}
for $m_{A^0} < 2 m_e$. The experiment is also sensitive to
$m_{A^0}$ up to 2.23 MeV, which is the $pn \rightarrow d\gamma$
transition energy. Requiring that the axion not decay before it
reaches the detector ($m_{A^0} C_{Aff} < 0.3 \MeV$), one
obtains the bulge at $m_{A^0} > 2 m_e$ in
Fig.~\ref{fig-ConstraintsReactor}.

%%%%%%%%%%%%%%%%%%%%%%%%%%%%%%%%%%%%%%%%%%%%%%%%%%%%%%%%%%%%%%%%%%%%%%%%%%%%%%%      beam dump limits      %%%%%%%%%%%%%%%

\subsection{Beam dump limits}

Axion--like particles can be emitted via bremsstrahlung or
Primakoff production in beam dump experiments (see,
e.g.~\cite{Bjorken:1988as}). The setup of these experiments is
as follows. An intense beam of particles (electrons or protons)
hits a thick target, which absorbs the beam and the interaction
products apart from very weakly interacting particles such as
axions. The decay products of the latter are collected by the
detector, typically placed tens of meters behind the target.

\begin{figure}[htb!]
\begin{center}
\includegraphics[width=9cm]{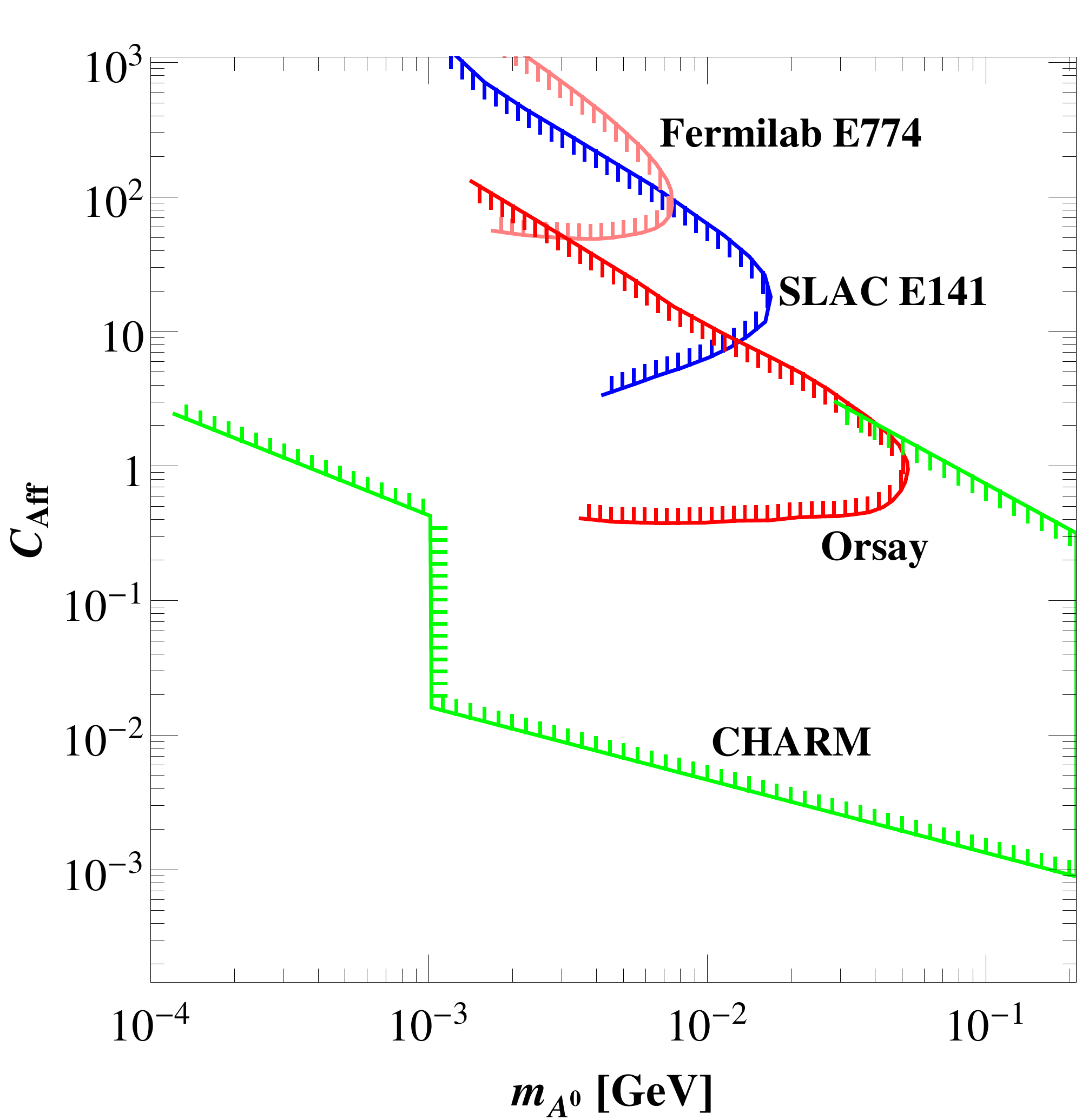}
\caption{Constraints from the beam dump experiments. \label{fig-ConstraintsBeamdump}}
\end{center}
\end{figure}

Below we consider 4 representative beam dump
experiments.\footnote{We are not displaying the results of the
SLAC E137~\cite{Bjorken:1988as} and Fermilab
605~\cite{Brown:1986xs} experiments since the corresponding
exclusion regions are largely covered by other experiments. }
\begin{itemize}
\item SLAC E141~\cite{Riordan:1987aw}: $2 \times 10^{15}$
    electrons  at energy 9 GeV struck a 12 cm tungsten
    target; detector 35 m behind the target
\item Fermilab E774~\cite{Bross:1989mp}: $0.52 \times
    10^{10}$ electrons at 275 GeV dumped at a 30 cm target;
    detector length 7.25 m
\item CHARM~\cite{Bergsma:1985qz}: $2.4 \times 10^{18}$
    protons at 400 GeV dumped at a thick copper target;
    detector 480 m behind the target
\item Orsay~\cite{Davier:1989wz}: $2 \times 10^{16}$
    electrons at energy 1.6 GeV dumped in a 1 m target;
    detector 2 m behind the target
\end{itemize}
The corresponding exclusion regions
(Fig.~\ref{fig-ConstraintsBeamdump}) can be read off from the
plots presented in these papers either using the conversion
factor for the axion decay constant (\ref{conversion}) or
calculating the axion decay time according to
Eq.~\ref{gamma-total}.

The reactor and beam dump experiments by themselves eliminate
most of the parameter space
(Figs.~\ref{fig-ConstraintsReactor},\ref{fig-ConstraintsBeamdump}).
They are based on a different kind of physics compared to meson
decays of Sec.~\ref{sec-lab} and in this sense are
complementary. Note also that, unlike meson decays, electron
beam dump experiments as well as the muon $g-2$ probe directly
the lepton--axion coupling, which could be the only axion
coupling to matter in exotic (``leptophilic'') scenarios.

%%%%%%%%%%%%%%%%%%%%%%%%%%%%%%%%%%%%%%%%%%%%%%%%%%%%%%%%%%%%%%%%%%%%%%%%%%%%%%%      Conclusions      %%%%%%%%%%%%%%%

\section{Conclusion}

In this work, we have studied the question how light a CP--odd
Higgs $A^0$ of the NMSSM can be. We have analyzed constraints
from meson decays, muon $g-2$, beam dump and reactor
experiments. We find that the parameter space $m_{A^0} < 2
m_\mu$ is excluded (by more than one experiment) unless the
coupling of $A^0$ to matter is 4 orders of magnitude smaller
than that of the Standard Model Higgs, i.e. $C_{Aff} <
10^{-4}$. Since such a small coupling can hardly be achieved in
the NMSSM, we conclude that $A^0$ has to be heavier than about
210 MeV.

Our analysis applies more generally to couplings of a light
pseudoscalar to matter. We have not used any specific features
of the NMSSM (nor supersymmetry). We have only relied on
Eq.~\ref{eq-L} and analyzed parameter space in terms of
\{$m_{A^0}, C_{Aff}$\}. Flavor changing couplings of the
Standard Model are sufficient to generate the processes like $B
\rightarrow K ~A^0$ and $K \rightarrow \pi ~A^0$, which lead to
strong constraints.

Since the CP--odd Higgs is heavier than $2 m_\mu$, the decay
channel to muons is open. One can therefore produce $A^0$ in
gluon fusion at the LHC and look for $\mu^+ \mu^-$ pairs with
low invariant mass~\cite{Dermisek:2009fd}, or search for decays
$h \rightarrow 2 A^0 \rightarrow 4
\mu$~\cite{Belyaev:2010ka,Goh:2008xz} (similar to what has been
done by D$0$ at Tevatron~\cite{Abazov:2009yi}). This will
provide an important test of models with a light pseudoscalar.

{\bf Acknowledgments.} A.R. would like to thank G. Weiglein for
useful discussions.

\bibliographystyle{utphys}
\bibliography{NMSSM-CPoddHiggs}
\addcontentsline{toc}{section}{References}

\end{document}